\documentclass[12pt]{article}
																																																																																											\pdfoutput=1																																																																																			 
\usepackage{amsmath,amssymb}
\usepackage{graphicx}
\usepackage[pdftex]{hyperref}
\DeclareGraphicsExtensions{.eps,.bmp,.wmf,.jpeg,.pdf}
\numberwithin{equation}{section}
\def\be{\begin{equation}}
\def\ee{\end{equation}}

\def\bea{\begin{eqnarray}}
\def\eea{\end{eqnarray}}
\title{Higgs Inflation with linear and quadratic curvature corrections}

\author{L. N. Granda\thanks{ngranda@univalle.edu.co}\, \ D.F. Jimenez \thanks{jimenez.diego@correounivalle.edu.co}\\ {\small\it Departamento de Fisica, Universidad del Valle}\\{\small\it A.A. 25360, Cali, Colombia}}
\date{}
\begin{document}
\maketitle

\begin{abstract}
\noindent We consider a single scalar field inflation model with Higgs potential and curvature corrections given by non-minimal derivative coupling to gravity and coupling to the Gauss-Bonnet invariant. Exact analytical expressions, within the slow-roll approximation, are obtained for the main physical quantities. These corrections lead to successful inflation driven by the $\phi^4$-potential with the main inflationary observables in the regions restricted by the latest Planck data. It is shown that these curvature corrections can make the $\phi^4$ potential not only compatible with the current CMB observations, but also consistent with the Standard Model Higgs phenomenology, achieving the possibility that the Higgs boson acts as the primordial inflaton. 

\end{abstract}

\section{Introduction}

\noindent So far the  theory of cosmic inflation is the most likely scenario for the eraly universe \cite{guth, linde, steinhardt}, since this theory accounts for the almost scale-invariant spectra of primordial density perturbations and provides the solution to the most pressing problems of the standard Big Bang cosmology, like the flatness, horizon, magnetic monopoles among others  (see \cite{revlinde, liddle, riotto, lyth, mukhanov, nojirio, nojirioo} for reviews). These primordial fluctuations grow to produce the observed large-scale structure and cosmic microwave background (CMB) temperature anisotropy \cite{planck18}. These results paved the way for all kinds of models that explain the cosmic inflation using the most appealing mechanism of slow-roll inflation, in which a sufficiently flat scalar field potential provides the necessary conditions for quasi-exponential expansion, in order to to meet the demands of inflationary phenomenology. The extreme energy and curvature conditions of the primordial universe causes problems like the initial singularity \cite{antoniadis, gasperini, kawai, cartier, Kanti:1998jd, toporensky}, quantum gravity noise, violation of the unitarity bound  \cite{burgess, espinosa}, that challenge all the inflationary models. The avoidance of these problems and the increasingly accuracy of the cosmological observations impose tight restrictions on the number of viable inflationary models. 
In addition to this, due to the high  curvature and energy involved during inflation, it has been hard to make contact with the scale of the high energy physics that is currently available to us. The well known and most studied chaotic-inflation models \cite{linde2} with quadratic and fourth-order potentials in the frame of the simplest single-scalar field model of inflation, have been discarded by the observational evidence mostly because they lead to a tensor-to-scalar ratio larger than the upper limit set by the latest data \cite{planck18, bkp}. Thus, the standard Higgs boson, which would be a fundamental candidate for inflation, is excluded if it is minimally coupled to gravity. Among the mechanisms used to lower the predictions for the tensor-to-scalar ratio, the most studied is the addition of non-minimal coupling to gravity term
of the type $\xi\phi^2R$ \cite{shaposhnikov}. But the successful inflation, in the case of Higgs boson non-minimally coupled to gravity, demands $\xi>>1$  which reaches the scale ($\sim M_p/\xi$) at which the unitarity bound of the theory is violated  \cite{burgess}, since $M_p/\xi$ could be close or below the energy scale of inflation. \\
Another approaches consider adding a term with derivatives of the scalar field coupled to the Einstein tensor with constant coupling function, i.e. $\frac{1}{2M^2}G_{\mu\nu}\partial^{\mu}\phi\partial^{\nu}\phi$, where $M$ is a constant with dimension of mass \cite{germani,  stsujikawa, nanyang}, and in \cite{granda3} a correction term of the form  $\frac{\xi}{2\phi^2}G_{\mu\nu}\partial^{\mu}\phi\partial^{\nu}\phi$ was proposed, where $\xi$ is a dimensionless constant which can take values in a  semi-infinite range \cite{grajim1}. This term also contributes to the solution of initial singularity problem \cite{foffa, cartier}.
In general terms, the role of this correction is to increase the gravitational friction that causes the scalar field to roll more slowly than with the sole influence of the potential. This term also allows successful inflation in small scalar field scenario. 
On the other hand, the simplest second-order curvature correction represented by the scalar field coupling to the Gauss-Bonnet term has been considered initially motivated by the solution of the initial singularity problem \cite{antoniadis, gasperini, kawai, cartier, Kanti:1998jd, toporensky}, and has been added to the simple scalar field model in order to obtain inflationary observables consistent with current data. In general terms, the GB correction may lead to suppression or increment of the tensor-to-scalar ratio, depending on the sign of the coupling and it also can increment the scalar spectral index towards the blue region, which imply tight constraints on the magnitude of the GB correction \cite{soda4, Guo:2009uk, Guo:2010jr, Jiang:2013gza, Kanti:2015pda, sergeioik}. The analysis of the combined effect of both, the non-minimal derivative and GB corrections in the dynamics of the slow-roll inflation driven by the potential has been proposed in \cite{grajim1, grajim2}.\\
In the present paper we propose the non-minimal derivative and GB curvature corrections to the action of the Higgs boson, and analyze their effect on the slow-roll dynamics, and the viability of slow-roll inflation driven by the Higgs potential. Although the slow-roll inflation is driven by the Higgs potential, the curvature corrections affect the scalar spectral index and the tensor-to-scalar ratio, among other observables, through the equation of motion of the scalar field. This approach can be considered as a hybrid between canonical and effective potential approach. Exact analytical expressions, within the slow-roll approximation, are obtained for the main physical quantities, and it is shown that the model leads to successful inflation conserving the Standard Model Higgs parameters without violation of quantum gravity bounds.
\section{The model and slow-roll equations}
\label{sec:2}
The model for the Higgs boson with curvature corrections given by its non-minimal derivative coupling to the Einstein's tensor and its coupling to the GB 4-dimensional invariant is 
\be\label{eqm1}
S=\int d^4x\sqrt{-g}\left[\frac{1}{2\kappa^2}R-\frac{1}{2}\partial_{\mu}\phi\partial^{\mu}\phi- \frac{\lambda}{4}\left(\phi^2-v^2\right)^2+F_1(\phi)G_{\mu\nu}\partial^{\mu}\phi\partial^{\nu}\phi-F_2(\phi){\cal G}\right]
\ee
where $\kappa^2=M_p^{-2}=8\pi G$, $G_{\mu\nu}$ is the Einstein's tensor, ${\cal G}$ is the GB 4-dimensional invariant given by
\be\label{eqm2}
{\cal G}=R^2-4R_{\mu\nu}R^{\mu\nu}+R_{\mu\nu\lambda\rho}R^{\mu\nu\lambda\rho}
\ee
where we consider a Higgs boson doublet of the form ${\cal H}=(0,\phi)$ with $\phi$ being a real scalar field. Note also that in what follows one can neglect the vev of the scalar field compared to its value during inflation. Here we consider inverse power-law coupling functions
\be\label{coupling functions}
F_1(\phi)=\frac{\xi}{\phi^2},\;\;\; F_2(\phi)=\frac{\eta}{\phi^4},
\ee
following the approach of (\cite{grajim1}). Power-law coupling for the GB correction has been proposed in \cite{ Guo:2010jr,  Jiang:2013gza, Koh:2014bka, sergeioik}, and in \cite{granda3} a power-law for the derivative coupling term has been proposed.
It is worth noticing that the derivative coupling can be interpreted as the coupling of the Ricci tensor to the free part of the matter energy-momentum tensor. In fact, the energy momentum tensor for the free scalar field is given by 
\be\label{freetensor}
T_{\mu\nu}=\partial_{\mu}\phi\partial_{\nu}\phi-\frac{1}{2}g_{\mu\nu}\partial_{\sigma}\phi\partial^{\sigma}\phi,
\ee
which coupled to $R_{\mu\nu}$ gives
\be\label{rcoupling}
R_{\mu\nu}T^{\mu\nu}=\left(R_{\mu\nu}-\frac{1}{2}g_{\mu\nu}R\right)\partial^{\mu}\phi\partial^{\nu}\phi=G_{\mu\nu}\partial^{\mu}\phi\partial^{\nu}\phi
\ee
An appealing property of the specific derivative coupling function $\xi/\phi^2$ is that the term
\be\label{kinetic}
\frac{\xi}{\phi^2}G_{\mu\nu}\partial^{\mu}\phi\partial^{\nu}\phi
\ee
has an interesting  symmetry: it is invariant under the transformation 
$$\phi\rightarrow \frac{1}{\phi}$$
which deserves further analysis. If this term were dominant (in the strong coupling limit this term gives consistent values for the inflationary observables as will be shown) in the Lagrangian, which is an allowed asymptotic case, then the physics for small and large field would be equivalent.\\
The field equations In the spatially flat FRW background 
\be\label{eqm4}
ds^2=-dt^2+a(t)^2\left(dx^2+dy^2+dz^2\right),
\ee
can written as follows 
\be\label{eqm5}
H^2=\frac{\kappa^2}{3}\left(\frac{1}{2}\dot{\phi}^2+\frac{\lambda}{4}\phi^4+9\frac{\xi}{\phi^2}H^2\dot{\phi}^2+24\eta H^3\frac{d}{dt}\left(\frac{1}{\phi^4}\right)\right)
\ee
\be\label{eqm6}
\dot{H}=\frac{\kappa^2}{2}\left[-\dot{\phi}^2+2(\dot{H}-3\frac{\xi}{\phi^2}H^2)\dot{\phi}^2+2\xi H\frac{d}{dt}\left(\frac{\dot{\phi}^2}{\phi^2}\right)+8\eta\frac{d}{dt}\left(H^2\frac{d}{dt}\left(\frac{1}{\phi^4}\right)\right)-8\eta H^3\frac{d}{dt}\left(\frac{1}{\phi^4}\right)\right]
\ee

\be\label{eqm7}
\ddot{\phi}+3H\dot{\phi}+\lambda\phi^3-96\frac{\eta}{\phi^5} H^2\left(H^2+\dot{H}\right)+6\xi H(3H^2+2\dot{H})\frac{\dot{\phi}}{\phi^2}+6\xi H^2\frac{\ddot{\phi}}{\phi^2}-6\xi H^2\frac{\dot{\phi}^2}{\phi^3}=0
\ee
In order to analyze we define the following slow-roll parameters 
\be\label{eqm8}
\epsilon_0=-\frac{\dot{H}}{H^2},\;\;\; \epsilon_1=\frac{\dot{\epsilon}_0}{H\epsilon_0}
\ee
\be\label{eqm10}
k_0=3\kappa^2F_1\dot{\phi}^2=3\xi\kappa^2\frac{\dot{\phi}^2}{\phi^2},\;\;\; k_1=\frac{\dot{k}_0}{Hk_0}
\ee
\be\label{eqm11}
\Delta_0=8\kappa^2H\dot{F_2}=-32\eta\kappa^2H\frac{\dot{\phi}}{\phi^5},\;\;\; \Delta_1=\frac{\dot{\Delta}_0}{H\Delta_0}
\ee
Using the slow-roll conditions $\ddot{\phi}<<3H\dot{\phi}$ and $k_0,\;... \;\Delta_1<<1$, we can reduce the field equations to obtain
\be\label{eqm14}
H^2\simeq \frac{\kappa^2}{3}V,
\ee
\be\label{eqm15}
\dot{H}\simeq \frac{\kappa^2}{2}\left(-\dot{\phi}^2-6H^2\frac{\dot{\phi}^2}{\phi^2}+32\eta H^3\frac{\dot{\phi}}{\phi^5}\right),
\ee
\be\label{eqm16}
3H\dot{\phi}+\lambda\phi^3+18\xi H^3\frac{\dot{\phi}}{\phi^2}-96\eta \frac{H^4}{\phi^5}\simeq 0,
\ee
The scalar field at the end of inflation is found from the condition $\epsilon_0(\phi_E)=1$, and the number of $e$-foldings before the end of inflation, using Eq. (\ref{eqm16}) is 
\be\label{eqm17}
N=\int_{\phi_I}^{\phi_E}\frac{H}{\dot{\phi}}d\phi=-\int_{\phi_I}^{\phi_E}\frac{H^2+6H^4F_1}{8H^4F'_2+\frac{1}{3}V'}d\phi=\int_{\phi_I}^{\phi_E}\frac{H^2\phi^5+6\xi H^4\phi^3}{32\eta H^4-\lambda\phi^8}d\phi
\ee
This equation allows to find the value of the scalar field at the horizon crossing, $\phi_E$, for a given number $N$ of $e$-folds. 

\subsection*{The Spectra of Scalar and Tensor Perturbations}
\noindent 
To find the second order action we use the uniform-field gauge and the perturbed metric around the FRW background given by
\be\label{pertmetric}
ds^2=-N^2dt^2+\gamma_{ij}\left(dx^{i}+N^{i}dt\right)\left(dx^{j}+N^{j}dt\right),
\ee
with
$$N=1+A,\;\;\; N^{i}=\partial^{i}B,\;\;\; \gamma_{ij}=a(t)^2e^{2\xi}\left(\delta_{ij}+h_{ij}+\frac{1}{2}h_{ik}h_{kj}\right)$$
where $A$, $B$ and $\xi$ are the the scalar metric perturbations, and $h_{ij}$ are tensor perturbations satisfying $h_{ii}=0$, $h_{ij}=h_{ji}$ and $\partial_ih_{ij}=0$. 
Expanding the action (\ref{eqm1}) up to second order in perturbations in the uniform-field gauge, we obtain the second order action for the scalar perturbations (see \cite{grajim1}) for details
\be\label{slr1}
\delta S_s^{2}=\int dt d^3xa^3{\cal G}_S\left[\dot{\xi}^2-\frac{c_S^2}{a^2}\left(\nabla\xi\right)^2\right]
\ee
where up to first order in slow-roll parameters
\be\label{slr2}
{\cal G}_S=M_p^2\left(\epsilon_0-\frac{1}{2}\Delta_0\right)
\ee
and 
\be\label{velocity}
c_S^2=1-\frac{4k_0\left(\epsilon_0+\Delta_0+\frac{4}{3}k_o\right)}{3\left(2\epsilon_0-\Delta_0\right)},
\ee
where $c_S$ is the velocity of scalar perturbations. This action leads to the following  $k$-dependence for the power spectra (see details in \cite{grajim1})
\be\label{slr28}
P_{\xi}=\frac{k^3}{2\pi^2}|\xi_k|^2\propto k^{3-2\mu_s},
\ee
and the scalar spectral index in terms of the slow-roll parameters  as
\be\label{slr29}
n_s-1=\frac{d\ln P_{\xi}}{d\ln k}=3-2\mu_s=-2\epsilon_0-\frac{2\epsilon_0\epsilon_1-\Delta_0\Delta_1}
{2\epsilon_0-\Delta_0}
\ee
The second order action for the tensor perturbations takes the form (\ref{grajim1})
\be\label{slrt1}
\delta S_2=\frac{1}{8}\int d^3xdt a^2{\cal G}_T\left[\left(\dot{h}_{ij}\right)^2-\frac{c_T^2}{a^2}\left(\nabla h_{ij}\right)^2\right],
\ee 
where in terms of the slow-roll parameters
 \be\label{gt}
 {\cal G}_T=M_p^2\left(1-\frac{1}{3}k_0-\Delta_0\right)
 \ee
 and 
 \be\label{tensorvelocity}
 c_T^2=\frac{3+k_0-3\Delta_0\left(\epsilon_0+\Delta_1\right)}{3-k_0-3\Delta_0}
 \ee
which leads to the power spectrum for tensor perturbations as
\be\label{slrt11}
P_T=\frac{k^3}{2\pi^2}|h^{(k)}_{ij}|^2
\ee
giving the following expression for the tensor spectral index, up to first order in slow-roll parameters
\be\label{slrt11}
n_T=3-2\mu_T=-2\epsilon_0
\ee
The relative contribution to the power spectra of tensor and scalar perturbations, defined as the tensor/scalar ratio $r$
\be\label{slrt12}
r=\frac{P_T(k)}{P_{\xi}(k)}.
\ee
For the scalar and tensor perturbations we can write the power spectra respectively as
\be\label{slrt13}
P_{\xi}=A_S\frac{H^2}{(2\pi)^2}\frac{{\cal G}_S^{1/2}}{{\cal F}_S^{3/2}},\;\;\; P_T= 16A_T\frac{H^2}{(2\pi)^2}\frac{{\cal G}_T^{1/2}}{{\cal F}_T^{3/2}}
\ee
where ${\cal F}_S=c_S^2{\cal G}_S$, ${\cal F}_T=c_T^2{\cal G}_T$ and 
\be\label{coeffa}
A_S=\frac{1}{2}2^{2\mu_s-3}\Big|\frac{\Gamma(\mu_s)}{\Gamma(3/2)}\Big|^2,\;\;\; A_T=\frac{1}{2}2^{2\mu_T-3}\Big|\frac{\Gamma(\mu_T)}{\Gamma(3/2)}\Big|^2,
\ee
and all magnitudes are evaluated at the moment of horizon exit when $c_s k=aH$.
Taking into account that $A_T/A_S\simeq 1$ when evaluated at the limit $\epsilon_0,\Delta_0,...<<1$, as follows from (\ref{slr29}) and (\ref{slrt11}), we can write the tensor/scalar ratio as follows
\be\label{slrt15}
r=16\frac{{\cal G}_T^{1/2}{\cal F}_S^{3/2}}{{\cal G}_S^{1/2}{\cal F}_T^{3/2}}=16\frac{c_S^3{\cal G}_S}{c_T^3{\cal G}_T}
\ee
which in terms of the slow-roll parameters takes the form
\be
\label{slrt16}
r=8\left(\frac{2\epsilon_0-\Delta_0}{1-\frac{1}{3}k_0-\Delta_0}\right)\simeq 8\left(2\epsilon_0-\Delta_0\right)
\ee
where the conditions during inflation $\epsilon_0,k_0,\Delta_0<<1$, allow to set  $c_T\simeq c_S\simeq 1$. 
It is worth noticing that this expression represents a modified consistency relation due to the GB coupling. We can write $r$ as
\be\label{slrt18}
r=-8n_T+\delta r=16\epsilon_0+\delta r,\;\;\; \delta r=-8\Delta_0,
\ee
where $r= -8n_T=16\epsilon_0$ is the standard consistency relation.

\section{The observational constraints}
\label{sec:3}
Here we place observational constraints on the model (\ref{eqm1}) with Higgs potential, where we neglect the vev $v$ of the scalar field compared to its value at the horizon crossing. Keeping the first order in slow-roll parameters, for all the main physical quantities can be given exact analytical expressions. This allows us to find reliable values for the observable quantities involved in inflation, like the scalar spectral index $n_s$ and the tensor-to-scalar ratio $r$. 
Using (\ref{eqm8})-(\ref{eqm11}), (\ref{eqm14})-(\ref{eqm16}) we find the following expressions for the slow-roll parameters in terms of the scalar field 

\be\nonumber
\epsilon_0=\frac{48M_p^4-32\lambda\eta}{3\lambda\beta\phi^4+6M_p^2\phi^2},\;\;\; \epsilon_1=\frac{32\left(3M_p^4-2\lambda\eta\right)\left(M_p^2+\lambda\beta\phi^2\right)}{3\phi^2\left(2M_p^2+\lambda\beta\phi^2\right)^2},\;\;\; \Delta_0=\frac{64\lambda\eta\left(3M_p^4-2\lambda\eta\right)}{9M_p^4\phi^2\left(2M_p^2+\lambda\beta\phi^2\right)},
\ee
\be\label{slowroll3}
\Delta_1=\frac{32\left(3M_p^4-2\lambda\eta\right)\left(M_p^2+\lambda\beta\phi^2\right)}{3\phi^2\left(2M_p^2+\lambda\beta\phi^2\right)^2},\;\;\; k_0=\frac{16\lambda\beta\left(3M_p^4-2\lambda\eta\right)^2}{9M_p^4\left(2M_p^2+\lambda\beta\phi^2\right)^2},\;\;\; 
k_1=\frac{32\lambda\beta\left(3M_p^4-2\lambda\eta\right)}{3\left(2M_p^2+\lambda\beta\phi^2\right)^2}.
\ee
Note that the coupling constant $\xi$ is dimensionless while $\eta$ has dimension of $mass^4$, and $\lambda$ always appears in products with the couplings $\beta$ and $\eta$. This makes more practical work with the dimensionless parameters $\alpha=\lambda\beta$ and $\gamma=(\lambda\eta)/M_p^4$. \\
From the condition $\epsilon(\phi_E)=1$ we find the scalar field at the end of inflation as
\be\label{phie3}
\phi_E=\frac{M_p}{\sqrt{\alpha}}\left[-1+\sqrt{1+16\alpha-\frac{32}{3}\alpha\gamma}\right]^{1/2}
\ee
Note that$\phi_E$ gets smaller as $\alpha$ increases. Integrating Eq. (\ref{eqm17}) we find the number of $e$-foldings as
\be\label{efolds}
N=\frac{3\phi^2\left(4M_p^2+\alpha\phi^2\right)}{M_p^4\left(64\gamma-96\right)}\Big|_{\phi_I}^{\phi_E}
\ee
which allows us to evaluate the scalar field at the horizon crossing, giving
\be\label{phiini}
\phi_I=\frac{M_p}{\sqrt{3\alpha}}\left[\sqrt{6}\sqrt{3+8(1+2N)(3-2\gamma)\alpha+\sqrt{9+48(3-2\gamma)\alpha}}-6\right]^{1/2}.
\ee
Using (\ref{slowroll3}) in (\ref{slr29}) and (\ref{slrt16}) we find  
\be\label{nsindex}
n_s=\frac{64\lambda\eta\left(2\lambda\beta(\phi_I/M_p)^2+3\right)+3\left(4(1-16\lambda\beta)(\phi_I/M_p)^2+4\lambda\beta(\phi_I/M_p)^4+\lambda^2\beta^2(\phi_I/M_p)^6-96\right)}{3(\phi_I/M_p)^2\left(\lambda\beta(\phi_I/M_p)^2+2\right)^2}
\ee

\be\label{scalartensor}
r=\frac{256\left(3-2\lambda\eta\right)^2}{9(\phi_I/M_p)^2\left(2+\lambda\beta(\phi_I/M_p)^2\right)},
\ee
where we replaced $\alpha=\lambda\beta$, $\gamma=\lambda\eta$ and $\phi_I$ is given by (\ref{phiini}). Note that at $\gamma_c=3/2$ the tensor-to-scalar ratio is canceled. It also follows from (\ref{scalartensor}) that the strong coupling regime for the GB interaction can lead to unphysical increment of $r$. In fact, for positive $\eta$ the range of $r$ is limited by $\gamma<\gamma_c$. For the kinetic coupling we have the following  behavior at the strong coupling limit
\be\label{stronglimit}
 \lim_{\beta \to \infty} n_s=\frac{2N-3}{2N+1},\;\;\;  \lim_{\beta \to \infty}r=\frac{16(3-2\lambda\eta)}{3(2N+1)}.
 \ee
 Thus, for $N=60$ the model predicts $n_s\simeq 0.967$ and for $\lambda\eta=1$ it gives $r\simeq 0.044$, which are within the region bounded by the Planck. In this limit a product $\lambda\eta\sim {\cal O}(1)$ give consistent values for $r$. This suggests that for large $\beta$, without reaching the strong coupling limit, and in the region $\lambda\eta\sim {\cal O}(1)$ we can get interesting physical results. In Fig. 1 we show some trajectories in the $(n_s,r)$-plane for three different initial conditions set by $N=50,\; 60,\; 70$.

\begin{figure}
\begin{center}
\includegraphics[scale=0.7]{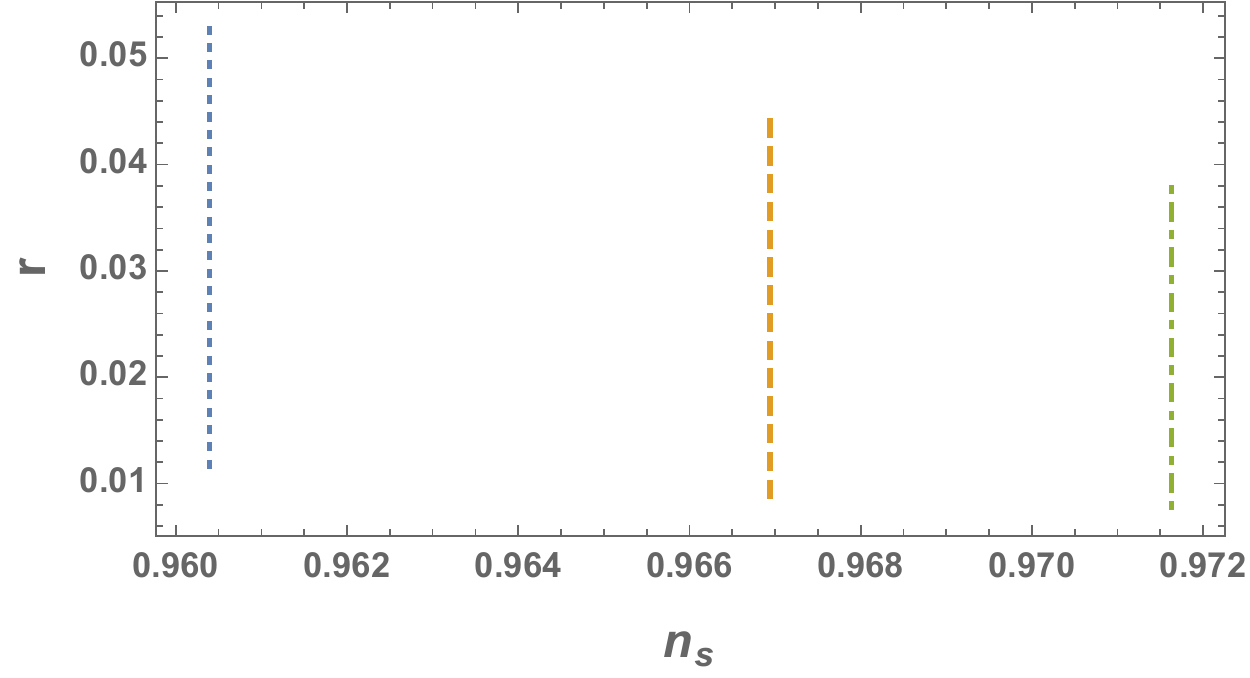}
\caption{The variation of the scalar spectral index $n_s$ and tensor-to-scalar ratio $r$ for $N=50$ (dotted), $N=60$ (Dashed) and $N=70$ (dot-dashed). We assumed $\alpha=10^{10}$ and $\gamma$ takes values in the interval $1\le \gamma\le 1.4$. Note that for a given $N$, the spectral index takes constant value, close to the strong coupling limit, while $r$ varies in an appropriate interval. The curves fall inside the region bounded by the latest observations.} 
\label{fig1}
\end{center}
\end{figure}

In Fig.2 we show the range of values of the scalar field at the horizon crossing and at the end of inflation, for the set of parameters considered in Fig. 1. 
\begin{figure}
\begin{center}
\includegraphics[scale=0.7]{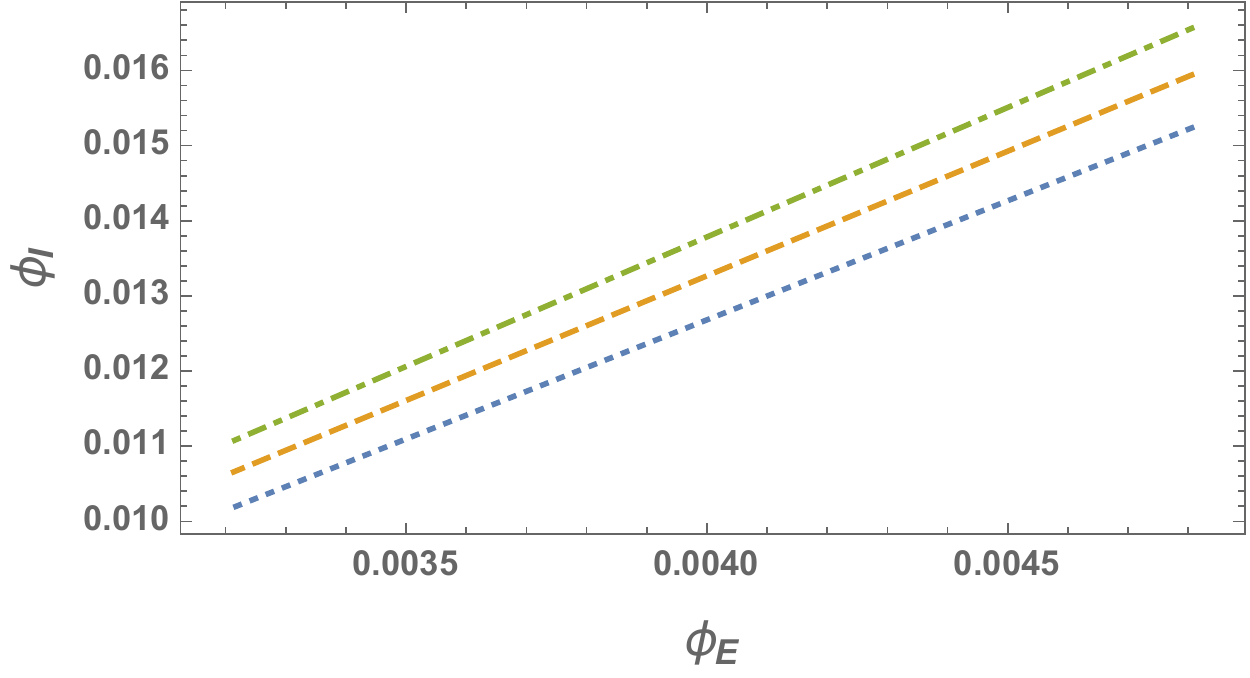}
\caption{The region of values taken by the scalar field at the horizon crossing versus its values at the end of inflation, for three different scenarios with $N=50$ (dotted), $N=60$ (Dashed) and $N=70$ (dot-dashed). As in Fig. 1 we assumed $\alpha=10^{10}$ and $\gamma$ in the interval $1\le \gamma\le 1.4$. The values of $\phi_I$ are consistent with the CMB normalization.} 
\label{fig2}
\end{center}
\end{figure}
In Fig. 3 we show the interval of values of the self-coupling $\lambda$ corresponding to the trajectories  $(n_s,r)$ shown in Fig. 1, for $\gamma<1.3$.
\begin{figure}
\begin{center}
\includegraphics[scale=0.7]{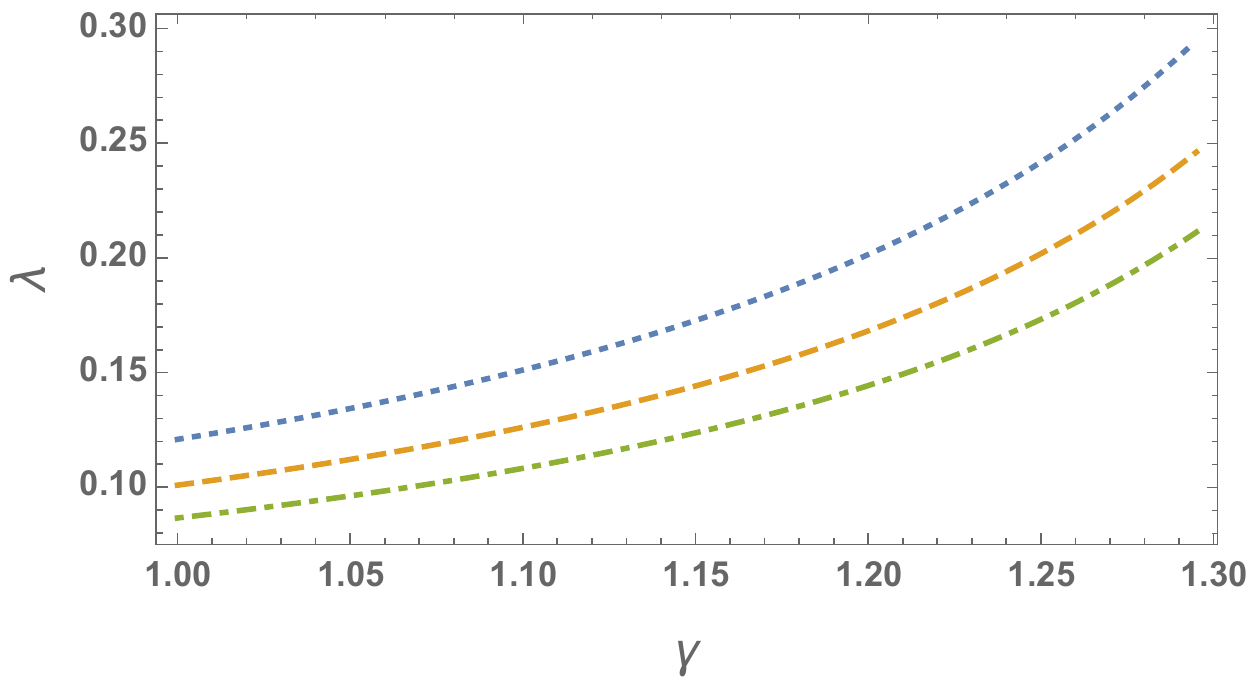}
\caption{The region of values taken by the self-coupling constant $\lambda$ for the three different scenarios with $N=50$ (dotted), $N=60$ (Dashed) and $N=70$ (dot-dashed). We assumed $\alpha=10^{10}$ and $\gamma$ in the interval $1\le \gamma\le 1.3$. The values of $\phi_I$ are consistent with the CMB normalization.} 
\label{fig3}
\end{center}
\end{figure}


\noindent To analyze the observational constraints imposed by CMB normalization, we observe that scale of the Hubble parameter during inflation can be evaluated using the COBE normalization for the power spectra, which can be written as  (\ref{slrt13}) 
\be\label{power-cobe}
P_{\xi}=A_S\frac{H^2}{(2\pi)^2}\frac{{\cal G}_S^{1/2}}{{\cal F}_S^{3/2}}\sim \frac{H^2}{2(2\pi)^2}\frac{1}{{\cal F}_S}\sim  \frac{H^2}{(2\pi)^2}\frac{1}{2\epsilon_0-\Delta_0}
\ee
where at the limit $(\epsilon_0,\epsilon_1,...)\rightarrow 0$ we can use the approximation $A_S\rightarrow 1/2$ and $c_S^2\rightarrow 1$. In order to show the consistency with COBE normalization and the Standard Model Higgs data, we take a sample of the parameters used in Fig. 1.\\
Taking $\alpha=10^{10}$, $\gamma=1$ and  $N=60$ we find from (\ref{slowroll3}) and (\ref{phiini}) the following values for the slow-roll parameters
\be\label{slow-rollini4}
\epsilon_0=0.0082,\;\; \epsilon_1=0.0165,\;\; \Delta_0=0.0110, \;\; \Delta_1=0.0165, \;\; k_0=0.0027,\;\; k_1=0.0165.
\ee
Form  (\ref{phiini}), (\ref{nsindex}) and (\ref{scalartensor}) we find
\be\label{phiininsr}
\phi_I=0.0159 M_p, \;\; n_s=0.967,\;\; r=0.044.
\ee
Then, replacing these values in (\ref{power-cobe}) we find
\be\label{cobeH4}
P_{\xi}\simeq 2.4\times 10^{-9}\sim  \frac{H^2}{(2\pi)^2}\frac{1}{0.005}\;\;\; \Rightarrow H\sim 2.2\times 10^{-5} M_p\sim 5.3 \times 10^{13} Gev.
\ee
On the other hand, from the tensor/scalar ratio we find
\be\label{cobescale4}
P_T=rP_{\xi}\sim 2\frac{H^2}{\pi^2M_p^2}\sim \frac{2V}{3\pi^2M_p^4}\sim (r) 2.4\times 10^{-9}\sim \;\;\;  \Rightarrow V\sim 1.63\times 10^{-9}M_p^4.
\ee
This constraint applied to the potential $V=\frac{\lambda}{4}\phi^4$ during inflation, using the value of $\phi_I$ from (\ref{phiininsr}), gives for the self-coupling constant $\lambda$ the value
\be\label{lambda}
\lambda\simeq 0.1
\ee
Hence this potential is consistent with CMB observations and the Standard Model Higgs parameters. From (\ref{lambda}) follows that $\beta\simeq 10^{11}$.\\
It is worth noticing that despite the fact that the kinetic coupling constant is large ( $\beta=10^{11}$), however the contribution of this term remains subdominant compared to the potential. We can evaluate the kinetic coupling to curvature term, from the Friedmann equation (\ref{eqm5}), during inflation as 
\be\label{kineticsub}
9\frac{\xi}{\phi^2}H^2\dot{\phi}^2\Big|_{\phi_I}=9\frac{\xi}{\phi_I^2}H^4\left(\frac{\partial\phi_I}{\partial N}\right)^2.
\ee
Taking the derivative of $\phi_I$ given in (\ref{phiini}) with respect to $N$ and using the parameters of the above numerical sample we find
$$\frac{\partial\phi_I}{\partial N}\simeq 6.6\times 10^{-5},$$
and replacing the values of $\phi_I$ and $H$ from (\ref{phiininsr}) and  (\ref{cobeH4}) it is found that 
\be\label{kineticscale}
9\frac{\xi}{\phi_I^2}H^4\left(\frac{\partial\phi_I}{\partial N}\right)^2\simeq 3.63\times 10^{-12}M_p^4<<V(\sim 1.63\times 10^{-9}M_p^4).
\ee
\section{Conclusions}
\label{sec:4}
In this paper we have studied the $\phi^4$-potential with curvature corrections given by the kinetic term non-minimally coupled to the Einstein tensor and the scalar field coupled to the GB invariant, and have verified that this field can realize successful inflation and, indeed, can be identified with the Standard Model Higgs boson.
The kinetic coupling term has the interesting property of being invariant under the transformation $\phi\rightarrow 1/\phi$, which deserves further study. 
For all considered scenarios with $N=50,\; 60,\; 70$, the model gives acceptable results consistent with the latest Planck data. 
Although the kinetic coupling constant takes large values ($\beta\sim 10^{11}$), however, the corresponding term in the Friedmann equation remains subdominant compared to the potential, as Eq. (\ref{kineticscale}) shows, which is consistent with the slow-roll dynamics driven by the potential. In addition to this, this value of the kinetic coupling constant
is enough to give $n_s$ very close to the strong coupling limit given in (\ref{stronglimit}), where $n_s=0.96,\; 0.967,\;  0.972$. for $N=50,\; 60,\; 70$ respectively. Indeed, evaluating $n_s$ from (\ref{nsindex}) with the data of Fig.1 for $\gamma=1$, it is found practically the above same values for $ns$. From Fig. 1 it follows that for a given $N$, the spectral index is almost constant along the considered $\gamma$-interval.
Moreover it should be noted that the spectral index remains practically the same along large $\beta$-interval (that covers various orders of magnitude), but to satisfy the COBE normalization and the Higgs boson restrictions, the coupling $\beta$ should be in the region $\beta\sim 10^{11}$. This behavior of $n_s$ being practically the same along the considered $\gamma$-interval and large $\beta$-interval, induces us to think that the model predicts $n_s$ and precisely gives fairly attractive value. The tensor-to-scalar ratio, which in the large kinetic coupling limit depends on $\eta$ (see (\ref{stronglimit})) takes also values in the interval bounded by the latest Planck observations. The behavior of the self-coupling $\lambda$ is depicted in Fig. 3 that shows the consistency with the Standard Model Higgs phenomenology.  As follows from (\ref{cobeH4}), the curvature scale involved in the inflation satisfies $R\simeq 12H^2<<M_p^2$, and therefore the quantum gravity bound is not exceeded. \\
The numerical analysis shows that the tensor-to-scalar ratio is controlled also by the allowed values of $\lambda$. According to the electroweak Standard Model, the allowed values of  $\lambda$ are in the range $0.1<\lambda\lesssim 0.27$ which bounds $r$ to the interval $0.017\le r\le 0.044$ for $N=60$. This result takes place independently of the large $\beta$ coupling limit (see (\ref{stronglimit})). Thus, in some sense the model also predicts the value of $r$ or at least gives tight interval for allowed values of $r$, which is well inside the region quoted by the latest observations.
We have proposed a framework in which the Higgs boson acts as the primordial inflaton. The considered curvature corrections to the inflationary scenario could help in the understanding of inflation from fundamental theories like the superstring theory.

\section*{Acknowledgments}
\noindent 
This work was supported by Universidad del Valle under projects CI 71195 and CI 71187. DFJ acknowledges support from COLCIENCIAS, Colombia.


\end{document}